\documentstyle[aas2pp4,epsf]{article}

\begin{document}

\title{The Ultramassive White Dwarf EUVE~J1746--706\footnote{Based on
observations obtained with NASA's {\it Extreme Ultraviolet Explorer}
and Mount Stromlo Observatory's 74 inch telescope.}}
\author{Jean Dupuis and St\'ephane Vennes}
\affil{Center for EUV Astrophysics\\
2150 Kittredge Street, University of California, Berkeley CA 94720--5030}
\authoremail{jdupuis@cea.berkeley.edu}

\vspace{1.5in}

\begin{abstract}

We have obtained new optical and extreme ultraviolet (EUV)
spectroscopy of the ultramassive white dwarf EUVE~J1746--706.  We
revise Vennes et al.'s (1996a, ApJ, 467, 784) original estimates of the
atmospheric parameters and we measure an effective temperature of
$46,500\pm700$ K and a surface gravity log~$g$ = $9.05\pm0.15$ ($\sim
1.2\,M_{\sun}$), in agreement with Balmer line profiles and the EUV
continuum. We derive an upper limit on the atmospheric abundance of
helium of He/H = $1.3 \times 10^{-4}$ and a neutral hydrogen column
density in the local interstellar medium $N_{\rm HI}\:=\:1.8 \pm 0.4
\times 10^{19}$ cm$^{-2}$ from the EUV spectrum. Our upper limit
corresponds to half the helium abundance observed in the atmosphere of
the ultramassive white dwarf GD~50. We discuss the possibility that
EUVE~J1746--706 represents an earlier phase of evolution relative to
GD~50 and may, therefore, help us understand the origin and evolution
of massive white dwarfs.

\end{abstract}

%\keywords{ISM: abundances --- stars: abundances --- stars: individual:\\
%EUVE~J1746--706 --- ultraviolet:stars --- white dwarfs}

\twocolumn

\section{INTRODUCTION}

Vennes et al. (1996a) identified EUVE~J1746--706 with a new
ultramassive white dwarf in extreme ultraviolet (EUV) all-sky surveys
({\it EUVE}, Bowyer et al.  1996; {\it ROSAT} Wide Field Camera, Pye et al. 1995).
Previously, this star remained without a precise classification.
Vennes et al. initially reported a mass of $1.43 \pm 0.06 M_{\sun}$
based on analysis of the Balmer line profiles making EUVE~J1746--706
one of the most massive white dwarfs known with a mass close to the
Chandrasekhar limit. Ultramassive white dwarfs are rare as shown by
various studies on the mass function of white dwarfs (see e.g.,
Koester, Schulz, \& Weidemann 1979; Bergeron, Saffer, \& Liebert 1992;
Bragaglia, Renzini, \& Bergeron 1995; and Finley 1995).  Among the few
previously known ultramassive single white dwarfs are GD~50 
($1.2 M_{\sun}$; Bergeron
et al. 1991), PG~0136+251 ($1.2 M_{\sun}$) and the magnetic white dwarf
PG~1658+441 ($1.3 M_{\sun}$; Schmidt et al. 1992). Since then, EUV surveys have
yielded new ultramassive white dwarfs (Vennes et al. 1996a; Vennes et
al. 1997) in larger proportion than the sample of stars selected in
the catalog of McCook \& Sion (1987) for the most recent white dwarf
mass distribution studies. A systematic study of this larger sample of
ultramassive white dwarfs will help achieve a better understanding of
their origin. If they indeed form a new class of white dwarf,
explaining their formation with just single star evolution would be
difficult.

Vennes, Bowyer, \& Dupuis (1996b) recently announced the discovery of
strong absorption lines of singly ionized helium in the EUV spectrum
of the ultramassive white dwarf GD~50.
The presence of helium in the atmosphere of a massive
white dwarf is paradoxical; Vennes et al. (1996b) speculate that it is
possibly the signature of helium dredge-up from the envelope by
meridional circulation currents.  Whether the same phenomenon could be
observed in other ultramassive white dwarfs or whether rapid rotation
is a characteristic of only a few massive white dwarfs remains
unclear. Bergeron et al. (1994) studied properties of a large sample
of DAO white dwarfs, but none of these objects bears any resemblance
to GD~50 or other massive white dwarfs, thus leaving the question open
as to the exact nature of these peculiarities. Optical and EUV
observations of EUVE~J1746--706 and similar objects may, therefore, help
establish ``class'' properties for massive white dwarfs, particularly
concerning the helium abundance pattern.

Since the initial report by Vennes et al. (1996a), we have secured
optical and EUV spectroscopy to improve our knowledge of the
fundamental parameters of this exceptional white dwarf. We present a
multiwavelength study of the new ultramassive white dwarf
EUVE~J1746--706, and we search, in particular, for the presence of
helium in the EUV spectrum. In \S~2, we describe the
new optical and EUV observations. In \S~3, we derive new values of the
effective temperature and surface gravity using both the Balmer line
profiles and the EUV energy distribution. We also
constrain the helium abundance with a multiparameter study of the
EUV continuum. In \S~4, we discuss EUVE~J1746--706 in
light of its possible origin as a merger of a double degenerate and
compare it to the case of GD~50. Finally, we summarize in \S~5.

\section{OBSERVATIONS AND DATA\newline REDUCTION}

\subsection{EUV Observations}

The EUV spectrum of EUVE~J1746--706 was acquired using NASA's {\it
Extreme Ultraviolet Explorer} ({\it EUVE\/})
between 1996 March 27 and April 5 with
an effective exposure time of about 180,000 s. The data were reduced
using the standard EUV package in IRAF with version 1.13 of the
calibration data described in the {\it EUVE} Guest Observer Software
User's Guide (Miller \& Abbott 1995). The star is detected in the
short wavelength spectrometer (SW, 70--190 \AA) with a maximum flux of
the order of $5 \times 10^{-4}$ photons~cm$^{-2}$~s$^{-1}$~\AA$^{-1}$
at 125 \AA\ and signal-to-noise ratio of 5 (Fig. 1).
Inspection of the images of the medium wavelength (MW, 140--380 \AA)
and long wavelength (LW, 280--760 \AA) spectrometers reveals no
significant spectra. The stellar emission is observed between 75 and 175 \AA\ and
is characterized by a featureless
continuum typical of the EUV spectrum of a DA star; the absence of
flux at long wavelengths is the result of attenuation in the
interstellar medium (ISM). We have also plotted the EUV spectrum of
GD~50 (Vennes, Bowyer, \& Dupuis 1996b), which is mentioned in the
introduction as being the sole case of a massive white dwarf showing
the series of \ion{He}{2} absorption lines in the EUV. Those lines are
not detected in the spectrum of EUVE~J1746--706 partly because the
hydrogen column density is much larger (see \S 3). However, this does
not exclude the presence of helium in the atmosphere: GD~50 would
appear similar to EUVE~J1746--706 if attenuated by a larger ISM column
density. We will exploit the sensitivity of the SW continuum to the
helium opacity in order to limit the the He abundance in the
photosphere of EUVE~J1746--706.

\begin{figure}[t]
\begin{center}
\hspace{0.1in}
\epsfxsize=0.47\textwidth
\epsfbox[50 200 541 666]{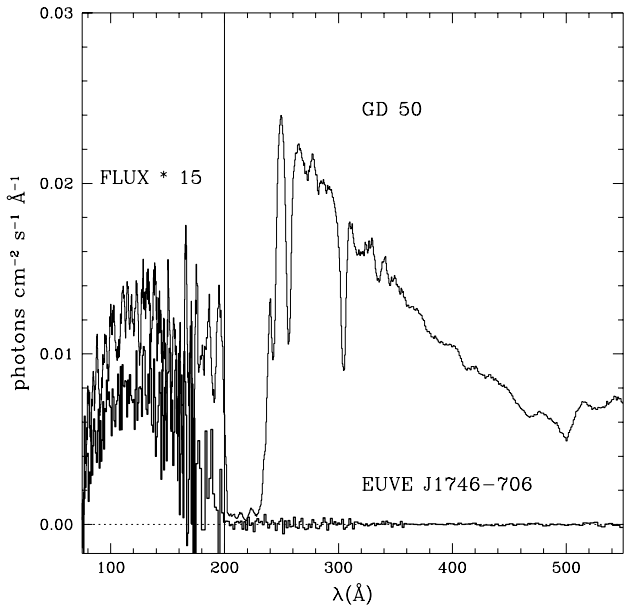}
%\plotone{ip31_fig1.ps}
\caption{Extreme ultraviolet spectroscopy of the ultramassive white
dwarf EUVE~J1746--706 obtained with the {\it Extreme Ultraviolet Explorer}
in 1996 April. The spectrum of GD~50 is also plotted ({\it thin line}) and is
detected throughout the EUV range (70--760 \AA) because of the low
column density of hydrogen in this line of sight ($\sim 2 \times
10^{18}$ cm$^{-2}$). Strong absorption lines of
photospheric helium are also present which would not be detected in
EUVE~J1746--706
because of its much larger hydrogen column density ($\sim 2 \times 10^{19}$
cm$^{-2}$). Both spectra are multiplied by 15 below 200 \AA.}
\end{center}
\end{figure}

%\vspace{-.15in}
\subsection{Optical Observations}

We reobserved EUVE~J1746--706 using the 74-inch telescope at Mount
Stromlo Observatory on 1996 April 25. We used the spectrometer at the
Cassegrain focus with a 300 lines~mm$^{-1}$ grating and a $1752
\times 532$ ultraviolet anti-reflection (UVAR) thinned SITE CCD; 
the images were binned $2 \times3$. Using standard IRAF procedures we
extracted a spectrum that covers the Balmer series between 3660 and
6040 \AA\ at 2.32 \AA\ per pixel or a spectral resolution of FWHM
$\approx$ 6 \AA. Figure 2 shows the optical spectrum of
EUVE~J1746--706, along with spectra of the massive white dwarf GD~50
and the DA white dwarf EUVE J1800+685 ($M = 0.5~M_{\sun}$; Vennes et
al. 1997); the two massive white dwarfs show distinctively broad
Balmer line profiles in comparison with EUVE~J1800+685. We adopted a
visual magnitude of $V=16.51$ (Bessell \& Dopita 1993), which allows
us to establish an absolute visual flux scale.

\begin{figure}[t]
\begin{center}
\hspace{0.1in}
\epsfxsize=0.47\textwidth
%\plotone{ip31_fig2.ps}
\epsfbox{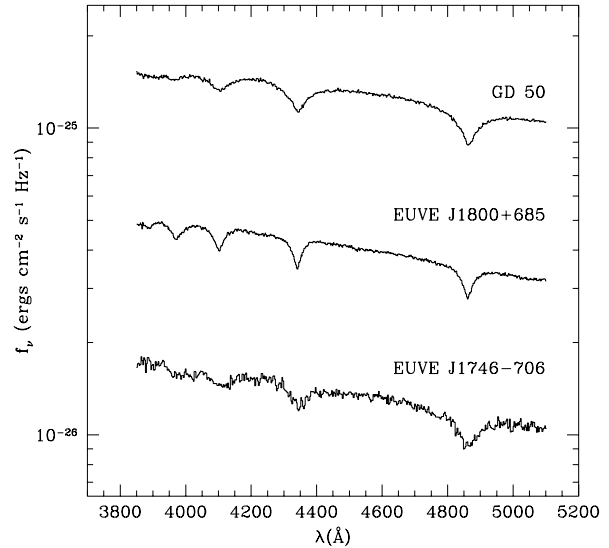}
\caption{Optical spectrum of EUVE~J1746--706, obtained with the 74-inch
telescope of Mount Stromlo Observatory in 1996 April, compared with
the spectra of GD~50 ($M \sim 1.2~M_{\sun}$) and EUVE J1800+685 ($M \sim
0.5~M_{\sun}$).
The Balmer lines in EUVE~J1746--706 have similar widths to those of
GD~50 but are much broader than those of EUVE~J1800+685 which has a
mass close to the average white dwarf mass ($\sim 0.6~M_{\sun}$).}
\end{center}
\end{figure}

\section{MODEL ATMOSPHERE ANALYSIS}

We measured the effective temperature and surface gravity by
simultaneously fittting the H$_\beta$, H$_\gamma$, H$_\delta$, and
H$_\epsilon$ line profiles with a grid of DA model atmospheres
computed with assumptions of local thermodynamic equilibrium and a
pure hydrogen composition. The same grid of models is described and
used by Vennes et al. (1996a); the grid points are between 20,000 K
and 32,000 K (in 2000 K steps) and between 32,000 K and 76,000 K (in
4000 K steps) for the effective temperature ($T_{\rm eff}$) and between
7.0 and 9.5 (in 0.5 steps) in the logarithm of the surface gravity
(log~$g$). A fit is obtained by means of a $\chi^2$ test between
the observed and synthetic line profiles convolved with the
appropriate spectral resolution. The right part of Figure 3 shows the
best solution along with the Balmer line spectra while the bottom left
corner of Figure 4 illustrates the 1, 2, and 3 $\sigma$ contours of
the fit (dashed line). Our new $T_{\rm eff}$ and log~$g$ are,
respectively, $46,200 \pm 2000$ K and $9.08 \pm 0.15$ and are only
marginally consistent with the values obtained by Vennes et al.
(1996a), $T_{\rm eff} = 41,300 \pm 1200$~K and log~$g = 9.40\pm0.12$.
Although our new surface gravity is lower than previously determined, it
still implies a membership in the class of ultramassive white dwarfs.
The difference between our value and previous results can only be 
explained in terms of different approaches in applying the flux calibration.

\begin{figure}[t]
\begin{center}
%\plotone{ip31_fig3.ps}
\plotfiddle{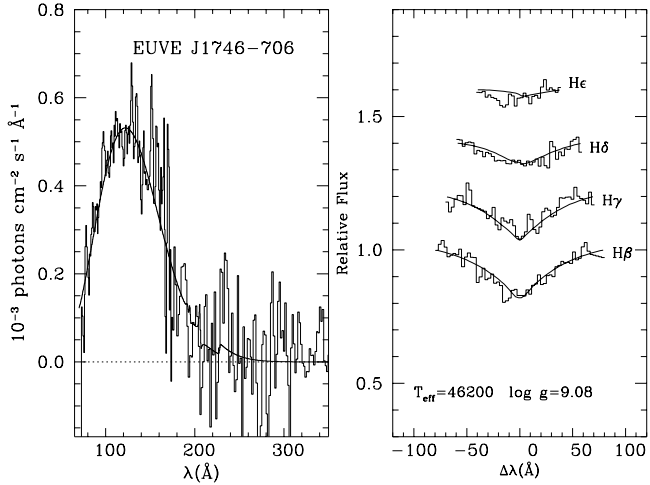}{2.2in}{-90}{40}{40}{-160pt}{204pt}
\vspace{0.2in}
\caption{Minimum $\chi^2$ fit of the EUV continuum and of the
Balmer lines of EUVE~J1746--706 using LTE DA white dwarf atmospheres
with homogeneous helium abundance. {\it Left}: We display a fit of the
EUV continuum (from 75--350 \AA) taking into account the effect of the
ISM absorption and of the presence of helium in the
photosphere.  {\it Right}: We present the solution of a simultaneous
fit of the Balmer line series (from H$_\beta$ to H$_\epsilon$) with
the assumption of a pure hydrogen composition.}
\end{center}
\end{figure}

\begin{figure}[t]
\begin{center}
\hspace{0.1in}
\epsfxsize=0.47\textwidth
\epsfbox[60 200 515 639]{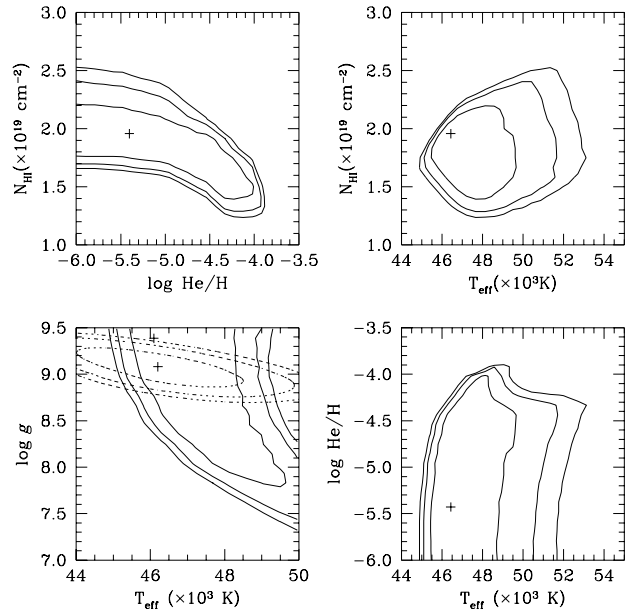}
% \plotone{ip31_fig4.ps}
\vspace{0.2in}
\caption{The 68.3\%, 90\%, and 99\% contours of
confidence in four different planes.
The crosses correspond to the location of
the $\chi^2$ minimum. The projected contours illustrate the
interdependence between the four free parameters: $T_{\rm eff}$, log~$g$,
log He/H, and log $N_{\rm H\,I}$. We also plot the contours for the Balmer lines
fit in the ($T_{\rm eff}$, $\log g$) plane ({\it dashed contours}).}
\end{center}
\end{figure}

The EUV spectrum is analyzed in a similar fashion but, in addition, we
constrain the abundance of helium in the atmosphere and the neutral
hydrogen column density in the local ISM\@.  We use a grid of LTE model
atmospheres computed with a homogeneous abundance of helium ranging
between log~He/H = --6 and --2 in steps of 0.5 dex.  (See Vennes et
al. 1996a for additional details about the grid and Vennes et al.
1996b for details concerning the validity of the LTE assumption in
massive white dwarfs.) The model spectra are normalized to the $V$
magnitude and convolved with the spectral resolution of the {\it EUVE}
Spectrometer ($\sim 0.5$ \AA\ in the SW, $\sim 1.0$ \AA\ in the MW,
$\sim 2.0$ \AA\ in the LW).  We fitted the flux-calibrated spectrum using
$\chi^2$ minimization and by varying $T_{\rm eff}$, log~$g$, He/H, and
$N_{\rm HI}$ (the ISM hydrogen column density). Because the continuum
is not detected in the MW and LW detectors, it is impossible to make
independent determinations of \ion{He}{1} and \ion{He}{2} column
densities; we set the ratio of the abundances of \ion{He}{1} and
\ion{He}{2} to \ion{H}{1} equal to 0.07 and 0.03 based on our
knowledge of other lines of sight in the local ISM (Vennes et al.
1994; Dupuis et al. 1995). The effective temperature is well
constrained at $46200 \pm 600$~K while the gravity is more loosely
constrained in the range between log~$g$ = 8.0 and 9.5.  We are able
to set a 3 $\sigma$ upper limit on the helium abundance of He/H $\leq
10^{-3.9}$, which is less than half the abundance of helium measured
in GD~50 (He/H $= 10^{-3.5}$; Vennes et al. 1996b). The neutral
hydrogen ISM column density is $1.8 \pm 0.4 \times 10^{19}$ cm$^{-2}$.
The best fit and the contour levels are shown in Figure 3 (left) 
and Figure 4, respectively.

We have superposed the Balmer lines contours to the EUV continuum
contours in the ($T_{\rm eff}$, log~$g$) plane in the lower left panel of
Figure 4. A unique solution is found by taking the intersection of
the 1 $\sigma$ contours, which further constrains the parameters to
$T_{\rm eff} =  46,500 \pm 700~{\rm K, \: and} \: \log g  =  9.05 \pm 0.15$.

Using the Hamada-Salpeter mass-radius relations for carbon interiors
(Hamada \& Salpeter 1961), we measure a mass of $1.21\pm0.07~M_{\sun}$
for EUVE~J1746--706. This value can be compared to $1.43\pm0.06~
M_{\sun}$ as previously determined by Vennes et al.'s (1996a)
extrapolation on Wood's (1995) mass-radius relations, or $1.32\pm0.03~
M_{\sun}$ when more appropriately interpolated on the Hamada-Salpeter
relations. Based on Hamada-Salpeter relations, the mass of GD~50 using
Bergeron et al.'s (1991) parameters is $1.19\pm0.07~M_{\sun}$. The
star EUVE~J1746--706 shares an identical mass with GD~50, but is
probably much younger.  Extrapolating on Wood's (1995) relations, the
cooling age of EUVE~J1746--706 is less than $\approx 50$ Myr, while
GD~50 is possibly $\approx 30$ Myr older.

\section{DISCUSSION}

Massive white dwarfs like EUVE~J1746--706 are rare. In their extensive
study of the white dwarf mass distribution, Bergeron, Saffer, \&
Liebert (1992) identified GD~50 as the only example of a massive
object. However, Vennes et al. (1996a) obtained a higher yield of
massive white dwarfs from a much smaller sample of 18 EUV-selected hot
white dwarfs. The origin of these stars is interesting from a stellar
evolution perspective because their masses exceed the upper limit of $
\sim 1.1~M_{\sun}$ for C/O white dwarfs, a limit set by the ignition
of non-degenerate core carbon burning (see reviews by Weidemann 1990
and Bragaglia, Renzini, \& Bergeron 1995).  Evolution of isolated
stars seems to produce white dwarfs with masses tightly concentrated
about $\sim 0.6~M_{\sun}$ with a possible tail extending toward
massive white dwarfs (Weidemann 1990; Bergeron, 
Saffer, \& Liebert 1992; Bragaglia,
Renzini, \& Bergeron 1995; Vennes et al. 1997). The progenitor of
EUVE~J1746--706 could be a massive star in view of the correlation
between the mass of a white dwarf and the mass of its progenitor
(Weidemann \& Koester 1983). However, estimating the mass of the
progenitor of EUVE~J1746--706 is difficult given the large
uncertainties in the initial-to-final mass relation. Nomoto (1984)
has suggested a scenario in which 8--10 $M_{\sun}$ stars could end
their evolution as O+Ne+Mg white dwarfs with masses in the range of
1.2--1.37 $M_{\sun}$. EUVE~J1746--706 and the other ultramassive white
dwarfs discovered by Vennes et al. (1996a,1997) could be examples of
O+Ne+Mg white dwarfs, although none of these objects are
expected to emerge from the local disk population.

Another scenario could be the merger of a
double-degenerate system as suggested by Bergeron et al. (1991) for
the case of GD~50. Yungelson et al. (1994) predict that 20\% of
merging double degenerates are CO-CO pairs and 30\% are CO-He pairs.
Are massive white dwarfs the results of CO-He mergers?
The {\it EUVE} spectrum of GD~50 supports this idea
because the helium abundance is much too high to be explained by a
process like radiative acceleration. Vennes et al. (1996b) discuss the
possibility that GD~50 is rapidly rotating, a residual of the
orbital angular momentum, and that helium is
transported to the surface by meridional circulation currents. Further
evidence in favor of a high rotational velocity comes from the flat
bottomed shape of the
\ion{He}{2} line series observed with {\it EUVE}.

Our results do not exclude a helium abundance half as high as in
GD~50. However, because a spectrum is only detected in the SW, we
cannot exclude the alternate possibility of a much lower helium
abundance.
Also, since the \ion{He}{2} lines are not detected
because of the high column density, no direct evidence exists of
rapid rotation other than a possible helium abundance comparable to
GD~50. Finally, a determination of the magnetic field strength may
allow a distinction between lineages with the massive DAp PG 1658+441
(Schmidt et al. 1992), and those with the non-magnetic massive DA GD~50
(Schmidt \& Smith 1995).

\section{CONCLUSIONS}

We have presented new spectroscopy of the ultramassive white dwarf
EUVE~J1746--706 in the optical and EUV range. We have reported a
unique solution for the atmospheric parameters that offers a good fit
to the Balmer lines and to the EUV continuum with $T_{\rm eff} = 46,500$~K
and log~$g = 9.05$. An upper limit on the helium abundance at about
half the measured abundance in GD~50 results from a modeling of the
EUV continuum.  At this point, we cannot exclude the possibility that
EUVE~J1746--706 is similar to GD~50 and that it may represent another
example of a merged double-degenerate system. A clear association of
GD~50 with EUVE~J1746--706 would emerge from a direct detection of
photospheric helium in the latter. A good approach would be
to look for the \ion{He}{2} $\lambda$4686 line in high-dispersion
spectra. An independent constraint on the neutral hydrogen column in
the local ISM would clarify the ambiguity between ISM attenuation and
photospheric helium absorption found in EUV data; a high-resolution
Hubble Space Telescope spectrum of the Ly$\alpha$ range may provide such a
constraint.

\acknowledgements

We would like to thank David Finley and Pierre Chayer for their useful
insights. This research is funded by the NASA grants NAG5--2620 and
NCC5--138 (JD). We gratefully acknowledge the editorial assistance of
Andrea Frank. The Center for EUV Astrophysics is a division of UC
Berkeley's Space Sciences Laboratory.

\end{document}